\documentclass[twocolumn,10pt]{article}
\usepackage[T1]{fontenc}
\usepackage[utf8]{inputenc}
\usepackage{mathptmx} 
\usepackage[scaled=.90]{helvet} 
\usepackage{courier} 

\usepackage[a4paper,top=0.6in,bottom=0.6in,left=0.6in,right=0.6in]{geometry}
\usepackage{graphicx}
\usepackage{xcolor}
\usepackage{cuted}
\usepackage{microtype}
\usepackage{parskip}
\usepackage{tcolorbox}
\tcbuselibrary{skins,breakable}
\usepackage{academicons}
\usepackage{hyperref}
\usepackage{lipsum}
\usepackage{lastpage}
\usepackage{booktabs}
\usepackage{caption}
\usepackage{enumitem}
\usepackage{fancyhdr}
\usepackage{titlesec}
\usepackage{float}
\usepackage[ruled,vlined]{algorithm2e}
\usepackage{wrapfig}
\usepackage{amsmath, amssymb}
\usepackage{multicol}
\usepackage{amsthm}

\usepackage[backend=biber,style=ieee]{biblatex}
\definecolor{primary}{HTML}{003049}      
\definecolor{accent}{HTML}{F77F00}       
\definecolor{textgray}{HTML}{2F2F2F}
\definecolor{dividergray}{HTML}{DADADA}

\hypersetup{
  colorlinks=true,
  linkcolor=primary,
  citecolor=primary,
  urlcolor=primary
}

\titleformat{\section}
  {\color{primary}\sffamily\Large\bfseries}
  {\thesection}{1em}{}[\vspace{0.2em}\titlerule]

\titleformat{\subsection}
  {\color{primary}\sffamily\large\bfseries}
  {\thesubsection}{1em}{}

\titlespacing*{\section}{0pt}{6pt}{3pt}
\titlespacing*{\subsection}{0pt}{4pt}{2pt}

\setlist[itemize]{left=1.2em, itemsep=2pt, topsep=2pt, parsep=0pt}
\setlist[enumerate]{left=1.2em, itemsep=2pt, topsep=2pt, parsep=0pt}

\tcbset{
  abstractstyle/.style={
    enhanced,
    colback=white,
    colframe=primary,
    boxrule=1pt,
    fonttitle=\bfseries\sffamily\large,
    left=6mm, right=6mm, top=2mm, bottom=2mm,
    width=\textwidth,
    boxsep=4pt,
    breakable
  },
  biobox/.style={
    colback=white,
    colframe=primary,
    arc=1.5mm,
    boxrule=0.5pt,
    drop shadow southeast,
    left=3mm, right=3mm, top=1mm, bottom=1mm,
    width=\linewidth,
    before skip=6pt, after skip=6pt,
    breakable
  }
}

\SetAlgoNlRelativeSize{-1}
\SetAlCapNameFnt{\sffamily\bfseries\small}
\SetAlCapFnt{\sffamily\small}

\makeatletter
\def\@maketitle{%
  \begin{center}
    {\fontsize{26pt}{20pt}\selectfont \bfseries \textcolor{primary}{The Hashed Fractal Key Recovery (HFKR) Problem: From Symbolic Path Inversion to Post-Quantum Cryptographic Keys}}\\[1ex]
    {\normalsize 
\textbf{
Mohamed Aly Bouke\,%
\href{https://orcid.org/0000-0003-3264-601X}{\includegraphics[height=1.8ex]{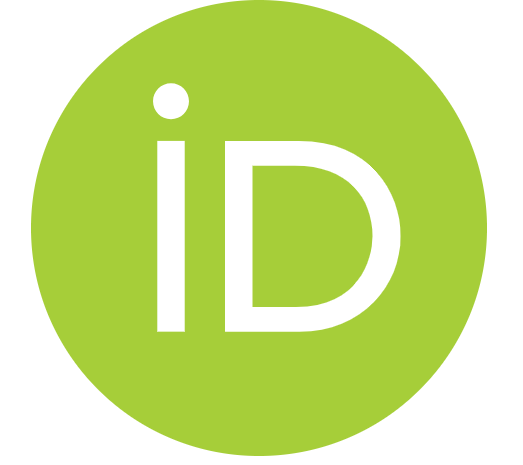}}%
\textsuperscript{1,*}
}
    }\\[0.8ex]
    {\footnotesize
      \textsuperscript{1}Department of Communication Technology and Network,\par Faculty of Computer Science and Information Technology, \par Universiti Putra Malaysia, Serdang 43400, Malaysia
    }\\[0.5ex]
    {\scriptsize \texttt{*bouke@ieee.org}}\\[1ex]
    {\scriptsize \textit{Researcher Paper} — \today}
  \end{center}
}
\renewcommand{\maketitle}{%
  \twocolumn[
    \color{textgray}
    \@maketitle
    \vspace{-1.2em}
  ]
}
\makeatother

\begin{document}
\maketitle

\begin{strip}
  \begin{center}
    \begin{tcolorbox}[abstractstyle, title=Abstract]
      
      \normalsize
        Classical cryptographic systems rely heavily on structured algebraic problems, such as factorization, discrete logarithms, or lattice-based assumptions, which are increasingly vulnerable to quantum attacks and structural cryptanalysis. In response, this work introduces the Hashed Fractal Key Recovery (HFKR) problem, a non-algebraic cryptographic construction grounded in symbolic dynamics and chaotic perturbations. HFKR builds on the Symbolic Path Inversion Problem (SPIP), leveraging symbolic trajectories generated via contractive affine maps over $\mathbb{Z}^2$, and compressing them into fixed-length cryptographic keys using hash-based obfuscation. A key contribution of this paper is the empirical confirmation that these symbolic paths exhibit fractal behavior, quantified via box-counting dimension, path geometry, and spatial density measures. The observed fractal dimension increases with trajectory length and stabilizes near 1.06, indicating symbolic self-similarity and space-filling complexity, both of which reinforce the entropy foundation of the scheme. Experimental results across 250 perturbation trials show that SHA3-512 and SHAKE256 amplify symbolic divergence effectively, achieving mean Hamming distances near 255, ideal bit-flip rates, and negligible entropy deviation. In contrast, BLAKE3 exhibits statistically uniform but weaker diffusion. These findings confirm that HFKR’s post-quantum security arises from the synergy between symbolic fractality and hash-based entropy amplification. The resulting construction offers a lightweight, structure-free foundation for secure key generation in adversarial settings without relying on algebraic hardness assumptions.

        \vspace{0.5em}

    \end{tcolorbox}
  \end{center}
\end{strip}

\section{Introduction} \label{sec:intro}
\vspace{0.8em}

Modern public-key cryptography is founded on hardness assumptions derived from structured algebraic problems such as integer factorization, discrete logarithms, and lattice-based constructions. These assumptions support widely adopted cryptosystems, including RSA, Diffie-Hellman, and more recent post-quantum candidates like Kyber and Dilithium. However, their reliance on algebraic structure introduces systemic weaknesses, particularly in light of quantum computing, which threatens many of these foundational problems \cite{rivest1984rsa,inam2025blockchain,barker2017recommendation}.

In response, researchers have begun to explore alternative approaches that do not depend on algebraic regularity. Symbolic dynamics and chaotic systems are emerging as promising candidates due to their inherent unpredictability, recursive complexity, and resistance to structured analysis \cite{morse1966symbolic,everett2024use}.

This work introduces the \emph{Hashed Fractal Key Recovery (HFKR)} problem, a cryptographic construction that combines symbolic chaos with hash-based entropy amplification. Building on the theoretical foundation of the Symbolic Path Inversion Problem (SPIP), HFKR uses contractive affine transformations over the integer lattice $\mathbb{Z}^2$ to generate symbolic trajectories, which are then compressed into fixed-length keys using cryptographic hash functions. This layered design avoids algebraic dependencies and leverages both the combinatorial explosion of symbolic paths and the diffusion properties of modern hash algorithms.

The main contributions of this paper are as follows: we formalize the HFKR problem as an extension of SPIP into a practical cryptographic setting, we empirically investigate the fractal properties and symbolic entropy of the generated trajectories, and we evaluate the ability of several hash functions SHA3-512, SHAKE256, and BLAKE3, to amplify symbolic divergence into secure cryptographic digests.

The remainder of this paper is organized as follows. Section~\ref{sec:related} surveys related work in chaos-based cryptography and symbolic dynamics. Section~\ref{sec:hkfr} defines the HFKR construction, including trajectory generation, hashing, and its fractal interpretation. Section~\ref{sec:empirical-HFKR} presents experimental results on symbolic entropy, diffusion metrics, and hash function performance. Finally, Section~\ref{conclusion} summarizes the contributions and outlines directions for future research.

\section{Related Work} \label{sec:related}
\vspace{0.8em}

Modern public-key cryptography is predominantly built upon algebraic hardness assumptions integer factorization, discrete logarithms, and lattice-based problems such as LWE and SVP. These problems underpin schemes like RSA, Diffie-Hellman, ECC, Kyber, and Dilithium \cite{bora2025synchronization,asif2021post,mikic2025post}. Their structured nature has enabled efficient implementations and strong theoretical guarantees. These schemes, while efficient and theoretically grounded, exhibit vulnerabilities under quantum algorithms such as Shor’s, which efficiently solves both integer factorization and discrete logarithm problems undermining the foundations of RSA and ECC.

This vulnerability has sparked interest in non-algebraic cryptographic constructions that do not rely on group theoretic or lattice-based regularities. Among such alternatives, chaotic systems and symbolic dynamics present a promising yet underexplored direction. These systems especially iterated function systems (IFS) and symbolic walks over discrete lattices offer non-reversible, high-entropy behaviors that resist algebraic simplification \cite{canto2023algorithmic,nejatollahi2019post,bouke2025rnifs}.

Prior uses of chaos in cryptography have largely focused on symmetric primitives, pseudorandom number generators, lightweight ciphers, and image encryption based on chaotic permutations. While these exploit the unpredictability of chaotic maps, they often lack rigorous complexity-theoretic underpinnings, limiting their suitability for asymmetric or public-key settings.

Some efforts have aimed to extend chaos into public-key cryptography. Mfungo et al. \cite{mfungo2023fractal} combined RSA with fractal geometry and chaotic maps to strengthen image encryption. Iovane et al. \cite{iovane2024quantum} proposed a quantum-resilient key generator based on fractals and stochastic prime sieving, albeit still grounded in RSA-like primitives. Al-Saidi et al. \cite{al2011efficiency} introduced a fractal public-key system using IFS, yet also relied on structured seeding via Diffie–Hellman.

While these designs incorporate chaos as an entropy source, most retain algebraic scaffolding or lack formal hardness analysis. A key limitation in the field remains the absence of structure-free, asymmetric cryptographic models with provable intractability.

The present work situates itself in contrast to these efforts by building upon a formally defined non-algebraic model, the Symbolic Path Inversion Problem (SPIP) \cite{bouke2025spip}. Rather than assuming the utility of chaos, SPIP rigorously proves that recovering symbolic trajectories over $\mathbb{Z}^2$ under bounded perturbations is both \#P-hard and PSPACE-hard. While SPIP deliberately refrains from proposing a full cryptographic scheme, it provides a solid theoretical foundation for non-invertible symbolic systems.

This paper extends that foundation into practical cryptography by introducing the HFKR problem a construction that leverages symbolic trajectories not only for theoretical hardness, but also for usable key generation via entropy amplification and hash-based obfuscation. In contrast to prior chaos-based approaches that often leap from heuristic design to application, HFKR is grounded in provable symbolic complexity, and advances a rigorously motivated, post-quantum-ready cryptographic primitive.

\section{The HFKR Problem} \label{sec:hkfr}
\vspace{0.8em}

The HFKR problem (Algorithm \ref{alg:HFKR}) introduces a novel cryptographic hardness assumption rooted in symbolic dynamics rather than algebraic structure. Unlike classical schemes that derive security from problems in number theory or lattice geometry, HFKR is built upon the combinatorial unpredictability of symbolic trajectories generated by chaotic affine systems. These trajectories evolve over the discrete lattice $\mathbb{Z}^2$, where each step is governed by a contractive, noisy affine transformation followed by discretization. While such systems naturally produce high entropy, their cryptographic utility depends on further transformation to mitigate any residual structure or statistical bias.

Formally, the symbolic path $\mathcal{P} = \{x_0, x_1, \dots, x_n\}$ evolves from an initial point $x_0 \in \mathbb{Z}^2$ under the recursive rule:
\begin{equation}
x_{i+1} = \left\lfloor A_i x_i + b_i + \delta_i \right\rfloor
\end{equation}
where each $A_i$ is a contractive linear map, $b_i$ is a bounded random translation, and $\delta_i$ is noise sampled from a bounded continuous distribution. The floor function enforces discretization, projecting the result back onto $\mathbb{Z}^2$. The resulting trajectory is non-invertible due to stochastic perturbations and rounding, and the number of possible symbolic walks grows exponentially as $|\mathcal{W}_n| = m^n$, where $m$ is the number of possible transformation configurations.

However, symbolic complexity alone does not guarantee cryptographic strength. Raw trajectories may leak partial information or exhibit statistical irregularities. To achieve cryptographic usability, HFKR appends a hash-based obfuscation layer:
\begin{equation}
k = H(x_0 \parallel x_1 \parallel \dots \parallel x_n)
\end{equation}
where $H$ is a cryptographic hash function such as SHA3-512. This step compresses the variable-length symbolic path into a fixed-size digest, amplifies symbolic differences via the avalanche effect, and enforces one-wayness. The result is a high-entropy key with no apparent structural correlation to its underlying symbolic source.

\begin{algorithm}[htbp]
\SetAlgoLined
\KwIn{
Initial point $x_0 \in \mathbb{Z}^2$; \\
Affine matrix parameters $(a_{11}, a_{12}, a_{21}, a_{22})$; \\
Translation bounds $[b_{\min}, b_{\max}]$; \\
Noise bound $\epsilon$; \\
Trajectory length $n$; \\
Hash function $H$
}
\KwOut{Cryptographic key $k$}
Initialize $\mathcal{P} \leftarrow [x_0]$\;

\For{$i \leftarrow 1$ \KwTo $n$}{
    Sample affine map $A_i$ and translation $b_i$\;
    Sample noise $\delta_i \sim \mathcal{U}([-\epsilon, \epsilon]^2)$\;
    Compute $x_i \leftarrow \lfloor A_i x_{i-1} + b_i + \delta_i \rfloor$\;
    Append $x_i$ to $\mathcal{P}$\;
}
Concatenate trajectory: $M \leftarrow x_0 \| x_1 \| \dots \| x_n$\;
Compute key: $k \leftarrow H(M)$\;
\Return $k$\;
\caption{HFKR Key Generation}
\label{alg:HFKR}
\end{algorithm}

While HFKR is described over a finite transformation set $\mathcal{T}$, our implementation adopts a stochastic extension, sampling fresh affine maps at each step. This maintains contractivity and unpredictability, as required by SPIP \cite{bouke2025spip}, while enhancing entropy through per-step variation. The resulting symbolic walk approximates a chaotic branching process with continuously evolving rules.

\subsection{Fractal Interpretation} \label{sec:fractal-link}
\vspace{0.8em}

Although SPIP operates in a discrete symbolic setting, its structural behavior closely mirrors that of classical fractal systems. The symbolic paths generated by contractive affine transformations with noise and discretization exhibit self-similar branching, exponential divergence, and non-reversibility all hallmarks of fractal geometry.

Symbolic fractality is not defined by visual self-similarity, but by recursive rule-based generation over a finite alphabet. As shown by Barnsley \cite{barnsley2014symbolic} and Kitchens \cite{kitchens2012symbolic}, symbolic substitution systems can produce fractal measures, topologies, and dimensions. Within this framework, the symbolic trajectory space $\mathcal{W}_n$ of HFKR behaves as a discrete fractal tree,  each step contracts the domain, branches stochastically, and introduces irreversibility via noise and rounding.

This behavior is empirically reflected in our experiments, where symbolic paths exhibit increasing coverage of the bounded state space with fractal dimension stabilizing near $D \approx 1.06$ (Section~\ref{sec:empirical-HFKR}). When hashed, these paths are compressed into fixed-length digests, projecting the symbolic fractal into a cryptographic output space. This mirrors the logic of fractal compression, but with irreversible mapping.

Thus, HFKR can be seen as a bridge between symbolic dynamics and fractal computation. It encodes a combinatorially rich symbolic structure into a secure digest, not by exploiting algebraic hardness, but by amplifying the entropy of symbolic fractals through cryptographic hashing. This fusion offers a new paradigm for post-quantum key generation, free from traditional algebraic assumptions.

\section{Empirical Evaluation}
\label{sec:empirical-HFKR}
\vspace{0.8em}

To investigate the extent to which the chaotic symbolic process in HFKR translates into measurable cryptographic unpredictability, we conducted a comparative experiment using three modern hash functions, SHA3-512, SHAKE256, and BLAKE3. Each function was applied to identical symbolic walks generated via the affine chaotic map described earlier, with carefully localized perturbations introduced mid-path. This setup isolates the effect of the hash function on the propagation of symbolic divergence into output-level entropy and bit-level diffusion.

\begin{figure}[htbp]
\centering
\includegraphics[width=\linewidth]{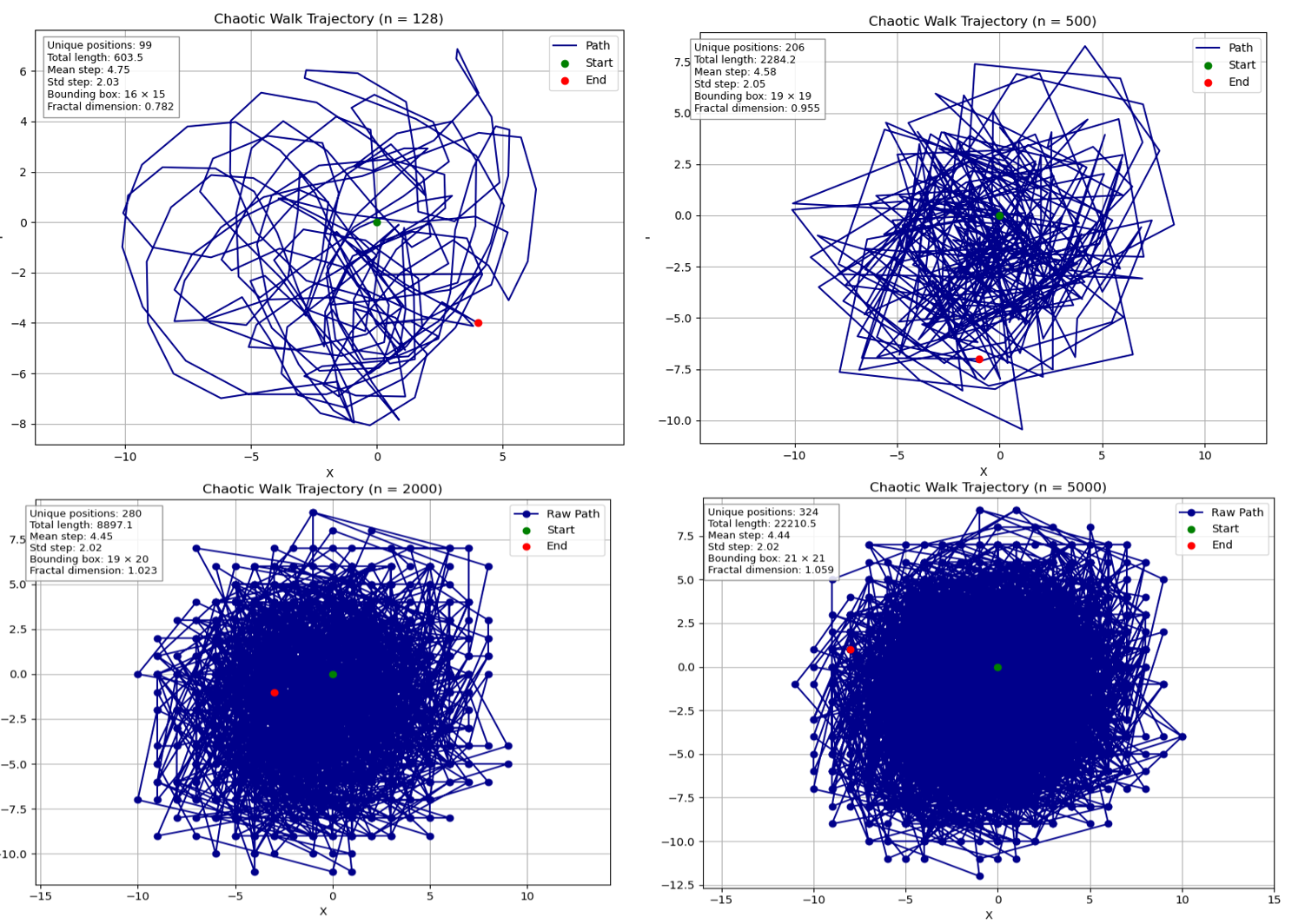}
\caption{Chaotic symbolic walks for increasing trajectory lengths ($n = 128, 500, 2000, 5000$). Geometric and fractal statistics are shown in each panel.}
\label{fig:walk}
\end{figure}

As illustrated in Figure~\ref{fig:walk}, the generated symbolic trajectories display progressively denser and more space-filling behavior as the number of steps $n$ increases. The paths exhibit characteristics commonly associated with chaotic dynamical systems, including recursive folding, sensitive dependence on initial conditions, and a lack of apparent periodicity. 

These properties are quantified through geometric and fractal metrics shown in each panel such as total path length, bounding box dimensions, and an estimated fractal dimension using the box-counting method. Notably, the fractal dimension increases with $n$, stabilizing near 1.06, which suggests that the symbolic trajectory increasingly occupies the surrounding space in a complex, self-overlapping manner. This behavior is consistent with symbolic fractality, making such walks highly suitable for entropy amplification and cryptographic use.

In the HFKR framework, these symbolic paths are flattened into byte streams and processed through cryptographic hash functions. This final transformation compresses their symbolic entropy into a fixed-length digest. The effectiveness of this entropy amplification is assessed through Hamming distance, bit-flip rate, and entropy deviation analyses.

\begin{figure}[htbp]
\centering
\includegraphics[width=\linewidth]{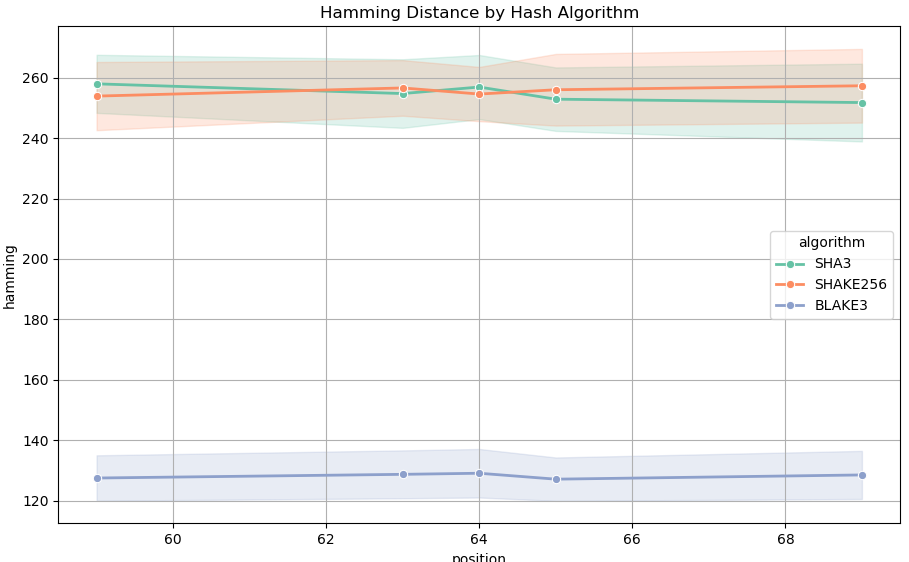}
\caption{Hamming distance by hash algorithm across perturbation positions.}
\label{fig:hamming}
\end{figure}

As shown in Figure~\ref{fig:hamming}, both SHA3-512 and SHAKE256 demonstrate strong avalanche properties, small symbolic perturbations result in high mean Hamming distances ($\approx$ 255 out of 512 bits). In contrast, BLAKE3 shows notably weaker diffusion, averaging only 128-bit divergence per perturbation. This suggests that BLAKE3's internal structure, which favors speed via Single Instruction, Multiple Data (SIMD) parallelism and tree-based hashing, may limit its capacity to fully amplify symbolic divergence.

\begin{figure}[htbp]
\centering
\includegraphics[width=\linewidth]{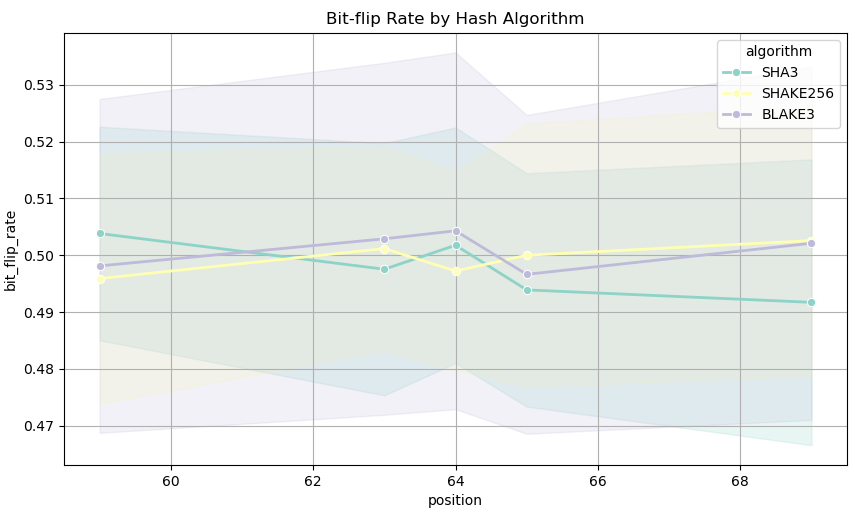}
\caption{Bit-flip rate across positions for different hash functions.}
\label{fig:bitflip}
\end{figure}

A similar story emerges in the bit-flip rate results shown in Figure~\ref{fig:bitflip}. While all three hash functions center around the ideal 0.5 bit-flip rate, SHA3 and SHAKE256 demonstrate tighter concentration and lower variance. BLAKE3, although still within acceptable cryptographic thresholds, shows broader fluctuations, which may indicate less uniform per-bit diffusion when applied to structured symbolic input.

\begin{figure}[htbp]
\centering
\includegraphics[width=\linewidth]{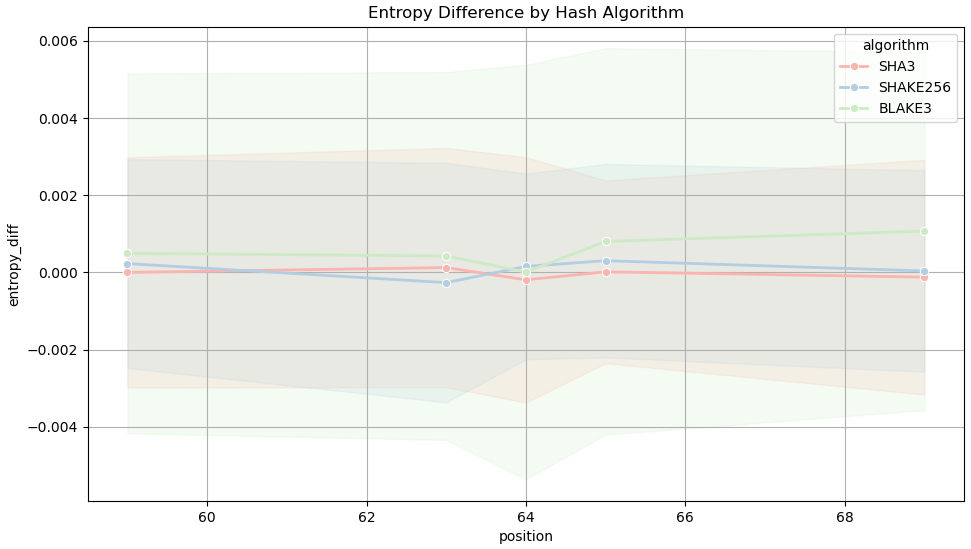}
\caption{Entropy difference between perturbed and original outputs.}
\label{fig:entropy}
\end{figure}

Entropy analysis, presented in Figure~\ref{fig:entropy}, confirms that the overall randomness of the outputs remains stable across all three functions. The entropy difference $\Delta H$ between perturbed and unperturbed hashes is effectively zero in aggregate, but again, BLAKE3 reveals a slightly broader envelope of variation. This suggests that while BLAKE3 maintains entropy levels, its internal mixing may respond less predictably to localized symbolic perturbations compared to Keccak-based constructions.

To further assess the positional uniformity of bit flips in the output digests, we applied a chi-square goodness-of-fit test to the output of each hash function. Each test was conducted on a cumulative bit matrix of size $250 \times 512$, corresponding to 50 independent trials across 5 perturbation offsets yielding 250 samples per output bit. This provides sufficient statistical depth to evaluate whether bit flips occur uniformly across positions, as would be expected under ideal avalanche conditions.

In conducting this test, each output bit is implicitly modeled as a Bernoulli variable with parameter $p = 0.5$, representing equal probability of flipping under symbolic perturbation. The null hypothesis asserts that flips are uniformly distributed across all bit positions, while the alternative hypothesis allows for any structured bias or diffusion weakness. The chi-square statistic thus measures deviation from this expected binomial behavior.

As summarized in Table~\ref{tab:chi}, all hash functions yielded p-values of $1.0000$, suggesting no statistically significant deviation from the uniform model. However, this result should not be mistaken for evidence of perfect or complete diffusion. With a large sample size, the test becomes highly sensitive, and when flip frequencies align closely with expectation, the test statistic can fall below rejection thresholds, producing p-values effectively rounded to $1.0000$. This reflects \emph{statistical alignment} with uniformity, but not necessarily \emph{cryptographic adequacy}.

Crucially, the chi-square test does not measure how strongly symbolic perturbations propagate across the output space. For example, both BLAKE3 and SHA3-512 achieve $p = 1.0000$, yet BLAKE3 exhibits a much lower mean Hamming distance and wider entropy variance. This indicates weaker symbolic amplification, despite statistical uniformity. In contrast, SHA3 and SHAKE256 show higher chi-square statistics and stronger diffusion effects, with more bits participating meaningfully in the output response.

These findings reinforce a core principle in HFKR, the hash function does not generate hardness ex nihilo, but acts as an amplifier of the symbolic entropy encoded by SPIP. Uniform bit-level activity ensures the absence of structural bias, but true security depends on how well this symbolic divergence is preserved and expanded across the digest space. The hash function must therefore be chosen not merely for cryptographic strength in isolation, but for its ability to faithfully propagate symbolic chaos into a secure output representation.

\vspace{0.6em}
\begin{table}[h]
\centering
\begin{tabular}{lccc}
\toprule
\textbf{Hash} & \textbf{Chi-square} & \textbf{p-value} & \textbf{Mean Hamming} \\
\midrule
SHA3-512      & 266.99              & 1.0000           & 254.85 \\
SHAKE256      & 257.09              & 1.0000           & 255.68 \\
BLAKE3        & 140.77              & 1.0000           & 128.21 \\
\bottomrule
\end{tabular}
\caption{Chi-square statistics and mean Hamming distances across 250 perturbation trials.}
\label{tab:chi}
\end{table}
\vspace{0.6em}

\section{Conclusion and Future Work} \label{conclusion}
\vspace{0.8em}

This paper introduced the HFKR problem as a non-algebraic cryptographic primitive grounded in symbolic dynamics and chaotic transformations. Building on the hardness foundation established by theSPIP, HFKR leverages symbolic trajectories generated by noisy contractive affine maps and transforms them into secure keys through cryptographic hashing. This layered construction departs from traditional algebraic assumptions, offering a structure-free model inherently resilient to quantum adversaries.

Empirical results confirm that symbolic paths produced in HFKR exhibit increasing fractal complexity and high entropy as their length grows. These trajectories demonstrate space-filling behavior and sensitivity to perturbations, with fractal dimensions converging near 1.06. When passed through hash functions, their symbolic divergence is amplified into robust cryptographic output. Specifically, SHA3-512 and SHAKE256 achieved strong avalanche properties maintaining high Hamming distances, near-ideal bit-flip rates, and consistent entropy, whereas BLAKE3 showed weaker symbolic diffusion despite statistical uniformity. These observations highlight that effective entropy amplification depends not only on the symbolic source but also on the diffusion strength of the hash function.

HFKR’s security arises from the synergy between symbolic unpredictability and irreversible hashing. Rather than compressing randomness passively, the hash function actively magnifies symbolic divergence into a one-way, high-entropy representation. This paradigm demonstrates that symbolic dynamics, when properly obfuscated, can serve as a viable foundation for post-quantum key generation.

Future work includes formalizing hardness reductions that link SPI style symbolic inversion with hash-based obfuscation under standard adversarial models. Extending HFKR into a full key encapsulation mechanism (KEM) and exploring its integration into hybrid protocols is also a promising direction. Moreover, evaluating its resilience under partial leakage or fault injection can provide insights into its real-world robustness. Given its lightweight, algebra-free nature, HFKR may be well-suited for deployment in constrained environments such as IoT devices or hardware enclaves. Lastly, optimizing symbolic encodings through grammar-based or fractal-aware schemes, could enhance both key density and implementation efficiency.

\section*{Declarations}
\vspace{0.8em}
\begin{itemize}
  \item \textbf{Funding:} Not applicable.
  
  \item \textbf{Conflict of Interest:} The authors declare that there is no conflict of interest.

\item \textbf{Availability of Data and Materials: } 
The full simulation code, along with configuration files and analysis scripts, is publicly available at:  
\href{https://github.com/drbouke/SPIP/blob/main/HFRK.py}{\texttt{https://github.com/drbouke/SPIP/}}.

\end{itemize}

\vspace{4.6em}

\section*{Author Biographies}
\begin{tcolorbox}[biobox]
  \begin{wrapfigure}{l}{0.35\linewidth}
    \vspace{-0.5em}
    \includegraphics[width=\linewidth]{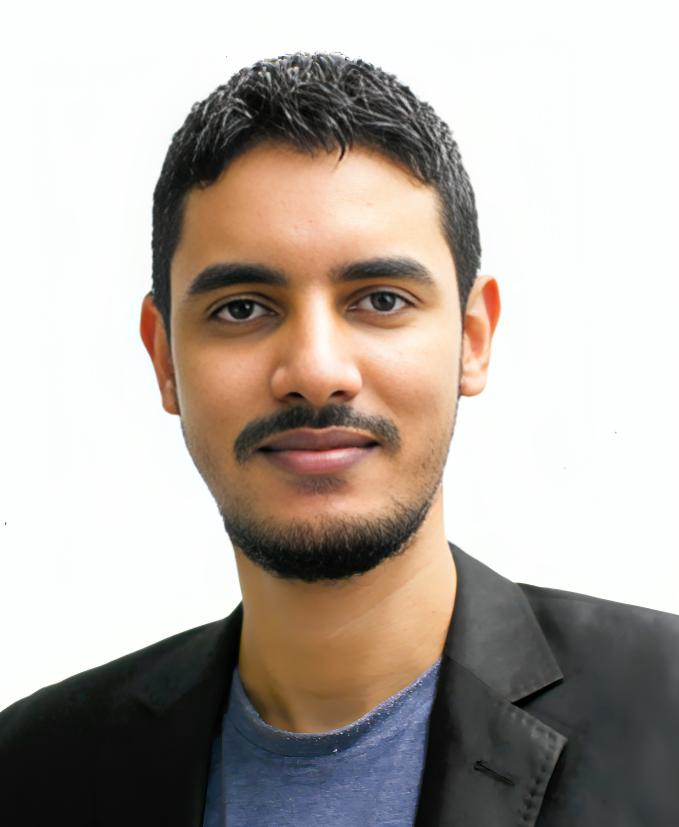}
  \end{wrapfigure}
  \textbf{Mohamed Aly Bouke}\,\href{https://orcid.org/0000-0003-3264-601X}{\includegraphics[height=1.8ex]{figs/orcid.png}} is a researcher with interdisciplinary expertise across theoretical mathematics, computer science, artificial intelligence, and cryptography. He holds a Master’s and a Ph.D. in Information Security from Universiti Putra Malaysia and has a background in mathematics education. His academic work spans topics such as mathematical modeling, epistemic systems, AI architectures, and secure computation. Dr. Bouke is an active member of the \textit{Institute of Electrical and Electronics Engineers (IEEE)}, the \textit{International Information System Security Certification Consortium (ISC2)}, and the \textit{Institute for Systems and Technologies of Information, Control and Communication (INSTICC)}. His contributions include peer-reviewed publications, invited talks, and academic training in both technical and theoretical domains.

  \vspace{0.8em}
  \textit{Email: bouke@ieee.org}
\end{tcolorbox}

\printbibliography

\end{document}